\documentclass[manuscript, nonacm]{acmart}

\usepackage{subcaption}

\AtBeginDocument{%
  }

\setcopyright{acmlicensed}
\copyrightyear{2018}
\acmYear{2018}
\acmDOI{XXXXXXX.XXXXXXX}
\acmConference[CHI '26]{Conference on Human Factors in Computing Systems}{April 23--28, 2026}{Barcelona Spain}
\acmISBN{978-1-4503-XXXX-X/2018/06}

\begin{document}

\title{Network Traffic as a Scalable Ethnographic Lens for Understanding University Students’ AI Tool Practices}


\author{Donghan Hu}
\affiliation{
  \institution{New York University}
  \city{NY}
  \state{NY}
  \country{USA}
}
\email{dh3915@nyu.edu}

\author{Rameen Mahmood}
\affiliation{
  \institution{New York University}
  \city{NY}
  \state{NY}
  \country{USA}
}
\email{rameen.mahmood@nyu.edu}

\author{Annabelle David}
\affiliation{
  \institution{New York University}
  \city{NY}
  \state{NY}
  \country{USA}
}
\email{ad7666@nyu.edu}

\author{Danny Yuxing Huang}
\affiliation{
  \institution{New York University}
  \city{NY}
  \state{NY}
  \country{USA}
}
\email{dhuang@nyu.edu }

\renewcommand{\shortauthors}{Anonymous et al.}

\begin{abstract}
AI-driven applications have become woven into students’ academic and creative workflows, influencing how they learn, write, and produce ideas. Gaining a nuanced understanding of these usage patterns is essential, yet conventional survey and interview methods remain limited by recall bias, self-presentation effects, and the underreporting of habitual behaviors. While ethnographic methods offer richer contextual insights, they often face challenges of scale and reproducibility. To bridge this gap, we introduce a privacy-conscious approach that repurposes VPN-based network traffic analysis as a scalable ethnographic technique for examining students’ real-world engagement with AI tools. By capturing anonymized metadata rather than content, this method enables fine-grained behavioral tracing while safeguarding personal information, thereby complementing self-report data. A three-week field deployment with university students reveals fragmented, short-duration interactions across multiple tools and devices, with intense bursts of activity coinciding with exam periods—patterns mirroring institutional rhythms of academic life. We conclude by discussing methodological, ethical, and empirical implications, positioning network traffic analysis as a promising avenue for large-scale digital ethnography on technology-in-practice.
  
\end{abstract}




\maketitle

\section{Introduction}

Ethnography has long been valued for its ability to generate “thick descriptions” of lived experience by situating human behavior within everyday contexts~\cite{geertz2008thick, lecompte1982problems, duque2019use, shapiro1994limits}. In fields such as education and HCI, ethnographic methods have illuminated how people integrate emerging technologies into routines that surveys or controlled experiments often miss~\cite{ejimabo2015effective}. Yet ethnography faces enduring challenges. Traditional fieldwork is expensive, labor-intensive, and time-consuming, often involving prolonged immersion with only a small number of participants. As a result, findings are rich but difficult to generalize. Moreover, ethnographic presence can alter participants’ behavior—the Hawthorne effect—making it challenging to capture authentic practices. Even rapid or lightweight ethnographies struggle with issues of scalability, evasiveness, and self-presentation, leaving gaps in understanding how new technologies are woven into daily life at the population scale.

The limitations of ethnography are particularly salient in the current wave of generative AI adoption in education practice. 
The release of widely accessible generative AI tools (Gen AI), particularly large language models (LLMs), has significantly transformed learning and education~\cite{openai_chatgpt, rodway2023impact}. Unlike earlier digital tools designed for information retrieval~\cite{google, smith2021using}, these systems generate novel human-like content, enabling students to draft text, summarize readings, and brainstorm ideas in ways that alter academic engagement~\cite{dempere2023impact, gasaymeh2025exploring, chen2025exploring}. Adoption has been rapid: in early 2023, ChatGPT became the fastest-growing consumer application, reaching more than 100 million users~\cite{acosta2024analysis}. Young adults, especially university students, comprise the majority of this user base~\cite{thormundson2023chatgpt}, positioning them as central actors in shaping how AI enters learning and daily life~\cite{ellis2023new}.

Research highlights both the promise and the tensions of this adoption. On the one hand, AI tools provide immediate feedback, encourage creative exploration, and support more personalized forms of learning~\cite{yim2025artificial, owan2023exploring, lo2023impact, vieriu2025impact, klimova2025exploring}. Students and educators increasingly recognize their value as tools for individualized assistance and cognitive extension~\cite{haroud2025generative, ali2023impact}. On the other hand, their unregulated and uneven use has raised concerns about academic integrity, algorithmic bias, data privacy, and the potential erosion of critical thinking~\cite{chan2023students, vieriu2025impact, al2024unveiling, ghimire2024guidelinesgovernancestudyai, larson2024critical}. These debates underscore why it is crucial to investigate how students actually use AI in everyday contexts, moving beyond attitudes and self-reports to capture authentic practices in situ. 
Current studies, however, rely heavily on self-reported data such as surveys, interviews, and diaries~\cite{gasaymeh2025exploring, balabdaoui2024survey, idroes2023student, haroud2025generative, ali2023impact, chan2023students}. While these methods capture perceptions, they are prone to recall errors, estimation mistakes, and social bias~\cite{zhang2025secret, haroud2025generative}. In academic contexts, where the line between legitimate assistance and misconduct is ambiguous~\cite{schiel2023high}, institutional pressures further discourage students from disclosing actual practices. As a result, self-reports often reflect idealized accounts rather than everyday behavior.

To bridge this gap, we propose treating digital network traffic as a form of trace ethnography \cite{geiger2011trace}: infrastructural data that captures cultural practices through the imprints they leave behind. By focusing on anonymized metadata such as domains, timestamps, and session volumes, researchers can unobtrusively observe how AI services are integrated into daily routines without capturing personal content. This approach preserves ethnography’s commitment to studying practices in situ, but it extends reach by offering breadth and continuity at scale. It also mitigates issues that arise when participants are conscious of being observed: because network logging operates passively—essentially a “set and forget” process—it reduces the reactivity associated with the Hawthorne effect~\cite{adair1984hawthorne}. Significantly, rather than replacing ethnography’s interpretive depth, it augments it with scalable, privacy-aware traces of daily practice, capturing aggregate patterns alongside facets of lived experience that direct observation may miss.

In this paper, we present the use of VPN-based network traffic analysis as a methodological extension of ethnography and apply it to the case of university students’ engagement with generative AI tools. AI use presents a salient case: adoption is rapid, practices are diverse and evolving, and behaviors are often hidden or downplayed in self-reports due to concerns about academic integrity. By analyzing traffic flows across widely used tools such as ChatGPT and Copilot, we uncover the rhythms of AI use—fragmented sessions, diurnal routines, and bursts of activity during periods of academic stress—that might otherwise go unnoticed. In doing so, we demonstrate how network traffic analysis can function as a scalable, privacy-aware form of digital ethnography, one that preserves contextual richness, extends observation to broader populations, and complements traditional methods by offering new perspectives on the lived experience of technology use.

In this paper, we answer the following research questions:
\begin{itemize}
    \item RQ1: How can anonymous network traffic analysis be leveraged as a lightweight digital ethnography method for observing students’ in-situ engagement with generative AI tools at scale?
    \item RQ2: What usage patterns characterize students’ interactions with AI tools in daily routines? 
    \item RQ3: What is the lived experience of participating in continuous VPN-based logging, and what do their responses reveal about trust and the negotiation of privacy in ethnographic research?
\end{itemize}

In addressing these questions, our study makes three contributions: 
(1) it introduces VPN-based network traffic logging as a scalable and privacy-aware extension of ethnography; 
(2) it demonstrates how AI use is patterned by institutional and cultural temporalities, such as exam periods and daily rhythms; and 
(3) it surfaces participants’ negotiations of trust, privacy, and informing the ethics and design of future in-the-wild studies.


\section{Related Work}
\subsection{The Ethnographic Tradition: Depth at Cost}
Ethnography, rooted in anthropology, is a qualitative method characterized by immersive first-hand interaction with participants in their everyday environments~\cite{goodson2011overview}. Through extended observation and in-depth interviews, ethnographers produce “thick descriptions” that illuminate the routines, cultural meanings, and material practices that shape human behavior. In HCI and educational research, ethnography has played a central role in revealing how people incorporate tools and technologies into their daily lives, insights that are often inaccessible through surveys or controlled experiments~\cite{orr2016talking}. In recent decades, ethnographic approaches have been widely adopted beyond anthropology, particularly in technology-focused research~\cite{blomberg2013reflections}. Ethnographic insights have informed the design of interactive systems by highlighting usability issues, uncovering tacit user practices, and surfacing the social and organizational contexts surrounding technology use~\cite{chamberlain2012research, palombo2020ethnographic}. This includes the use of ethnomethodological approaches~\cite{crabtree2004taking} and contextual investigation to support human-centered design, especially in workplace settings~\cite{crabtree2009ethnography, hashizume2013understanding}.

In the education field, ethnographic research has similarly been instrumental in uncovering how learners engage with digital tools in everyday contexts. Educational ethnographers often investigate how learners’ day-to-day environments and cultural practices shape their engagement with tools and content. For example, Moll et al. conducted ethnographic home visits to uncover the “funds of knowledge”, the practical skills and knowledge resources in working-class families’ daily lives ~\cite{moll1992bilingual}. Their study revealed that households engaged in rich, habitual practices (e.g., mechanics, commerce, childcare) could be leveraged to improve classroom learning. More recently, Chen et al. conducted an ethnographic case study of young English learners using mobile devices at home, adopting a technology-based approach
~\cite{chen2021beyond}. By conducting in-depth interviews and participant observations, they examined how children’s mobile-assisted language learning was embedded in everyday life. Such findings underscore how ethnography in education uncovers context-specific habits and barriers. In addition, arising concerns about the cost of time, many of the observed appeals to ethnography in existing work have been driven by ‘rapid’ ethnography, where time spent within the settings being studied is exceptionally short~\cite{millen2000rapid}.

However, depth comes with trade-offs. Traditional ethnography demands significant time, labor, and resources, making it difficult to scale or generalize findings~\cite{forsythe1999s, hammersley2006ethnography}. Prolonged fieldwork—often spanning months or years—is a hallmark of this approach, resulting in studies that typically focus on small samples or single communities~\cite{goodson2011overview}. Unlike surveys that can collect responses from hundreds of participants, ethnographic research often involves intensive engagement with only a few individuals. This limits its ability to produce findings that extend to broader populations. In addition, the data produced through ethnography are often complex and unstructured, requiring substantial effort to interpret~\cite{hammersley2006ethnography}. Researchers must also account for subjectivity, critically reflecting on how their presence and perspective shape the data and its analysis~\cite{hegelund2005objectivity}.

In this work, our objective is to retain the contextual benefits of the ethnographic approach while addressing its practical limitations. To do so, we introduce a form of digital ethnography that leverages anonymized network traffic data to capture users’ real-world interactions with generative AI tools unobtrusively. By analyzing patterns such as usage frequency, tool-switching, and time of engagement, this method provides scalable and fine-grained insights into everyday AI tool use. While it does not replace in-person observation, it offers a lower-cost and privacy-preserving alternative for studying habitual behavior across diverse contexts and over time—extending ethnographic inquiry into a new methodological space.

\subsection{AI-Tools Usage Studies}
The growing adoption of AI tools among students is reshaping both academic practices and everyday routines. Recent surveys indicate high levels of awareness and widespread use of generative AI platforms among college students~\cite{stohr2024perceptions}. In higher education, tools such as ChatGPT have quickly become central to academic support, while applications such as Grammarly and code generation platforms are increasingly embedded in students’ digital habits. Importantly, this engagement is not limited to coursework. AI tools are also being integrated into students’ personal lives, further blurring the line between academic responsibilities and everyday activities~\cite{hwang2023review, nemorin2023ai}.

Emerging quantitative research has begun to document how frequently and in what ways students engage with these tools~\cite{acosta2024analysis, al2024unveiling, vieriu2025impact, gasaymeh2025exploring, chan2023students}. Usage patterns vary between demographics and institutional contexts. For example, a large US-based survey study involving over 1,000 participants found that more than half of the students used ChatGPT regularly for writing and general tasks~\cite{baek2024chatgpt}. Interestingly, older students in their thirties and forties were more likely to use the tool frequently compared to younger age groups, and non-native English speakers were particularly inclined to rely on it for language support~\cite{baek2024chatgpt}. Another survey-based study reported that students perceive AI tools as beneficial for personalized learning, academic performance, and general engagement~\cite{vieriu2025impact}. Community policies could also influence adoption: students attending universities that allow AI usage are more likely to integrate these tools into their work, while others may avoid or hide their use due to concerns about academic integrity policies~\cite{zhang2025secret}. These findings highlight how student backgrounds and specific institutional environments can shape their engagement with AI.

Understanding when and why students use these tools offers further insight into evolving academic practices. A global survey of over 23,000 higher education students from 109 countries found that students most frequently used ChatGPT for practical academic tasks, such as brainstorming ideas, summarizing texts, and identifying references~\cite{ravvselj2025higher}. In this work, many participants reported that the tool helped simplify complex concepts and reduce the time spent doing routine work, which in turn improved their productivity. In another study, students in a software engineering course were permitted to freely use AI tools during a term project~\cite{clift2025learning}. Every student did so, and the resulting projects were more comprehensive and technically ambitious than those from previous cohorts. Students attributed this improvement to the assistance AI provided in navigating complex programming challenges. These examples suggest that students use AI tools strategically, combining systems for tasks such as drafting, coding, and editing to improve efficiency and output quality.

Taken together, these studies highlight the importance of understanding how students engage with AI tools. Patterns of behaviors, underlying motivations, and interactive usage can influence academic practices and inform the future of education. Developing this comprehensive understanding requires attention not only to what tools students use, but also to when, how, and why they use them. Our study addresses this need by examining how VPN-based network traffic logging can capture such patterns at scale, what rhythms and practices characterize students’ use of AI in their daily routines, and how participants themselves experience this form of monitoring.

\subsection{The Self-Report Paradox}
Self-report methods, such as surveys, interviews, and diary studies, are widely used in HCI and education research to examine how individuals engage with tools, applications, and technologies in their everyday lives~\cite{hektner2007experience, paulhus2007self}. These approaches are favored for their ease of deployment, scalability, and ability to capture subjective experiences at a relatively low cost. For example, time-use diaries have long been employed to record daily activities, requiring participants to document their own behaviors in situ~\cite{van2010advances}. Surveys, in particular, continue to play an essential role in research on digital media and technology use~\cite{mahalingham2023assessing}, reflecting their accessibility and broad applicability across contexts. However, this methodological convenience comes with limitations that challenge the accuracy and validity of self-reported data.

A growing body of evidence highlights the vulnerability of self-report instruments to well-documented forms of bias. Routine and habitual digital behaviors—such as checking phones, switching between apps, or browsing AI tools—are difficult to recall precisely, which introduces recall bias. In parallel, social desirability bias may cause participants to misrepresent their behaviors in ways that align with perceived norms or expectations~\cite{burnell2021associations, wu2022overestimating}. As a result, discrepancies between self-reported and actual usage are common. Multiple studies have shown weak or inconsistent correlations between self-estimated behavior and objective usage logs~\cite{mahalingham2023assessing, burnell2021associations}. Users often misjudge how often or how long they use particular technologies: some overreport their engagement, while others consistently underreport it~\cite{wu2022overestimating}.

These inaccuracies are not evenly distributed. For instance, heavier users of a technology tend to produce less reliable estimates than lighter users~\cite{sewall2020psychosocial}, and certain demographic groups, such as male participants, have been found to systematically overestimate usage~\cite{sewall2020psychosocial}. This pattern introduces noise and potential error into research findings, especially when exploring links between digital behavior and learning outcomes. Studies of adolescent social media use, for example, have demonstrated that reliance on self-reported usage data can lead to inflated or misleading conclusions~\cite{boyle2022systematic}. In response, researchers increasingly advocate for integrating objective behavioral measures—such as device logs, app tracking, or passive sensing—to complement or replace self-reports and improve the reliability of findings~\cite{burnell2021associations, wu2022overestimating}.

Addressing these gaps, our study examines how anonymized network traffic logging can capture patterns of student engagement with AI tools that self-report methods often miss. Specifically, we investigate not only how frequently students use these tools, but also when usage occurs, how it shifts over time, and how participants themselves experience being monitored. In doing so, we position network-based logging as a bridge between the contextual depth of ethnographic approaches and the scalability needed to study everyday AI use.

\section{Method}

We approached this study as an exploration of how VPN-based network traffic monitoring can serve as a digital ethnographic method for capturing everyday use of AI tools. A Virtual Private Network (VPN) is typically used to secure online activity by routing traffic through an encrypted server~\cite{ferguson1998vpn}. In this study, we repurposed this infrastructure: rather than building custom monitoring software, we used the VPN to log participants’ outgoing requests in a secure, privacy-preserving way. This setup captured when and how students accessed AI-powered applications—without recording message content or personal files. The approach complements self-report methods by producing continuous digital traces that situate participants’ AI use in the flow of daily life.

We designed a three-week field deployment combining passive VPN logging with surveys and interviews. This mixed-methods approach allowed us to examine the feasibility of VPN-based monitoring and participants’ lived experiences of being studied in this manner, and how such an approach might enrich and complement more traditional ethnographic methods.

\subsection{Recruitment \& Participants}
We recruited participants from a graduate-level engineering course, where the instructor shared our invitation to join the study. Interested students completed an initial sign-up form and received an information packet detailing the study procedures, including plain-language explanations of what data would be logged, associated risks, and safeguards. Enrollment took place through a secure portal, where participants reviewed and signed the consent form electronically, generated their own participant identifier (PID), and configured the monitoring system. Participation was voluntary: students could decline, withdraw, or regenerate their PID at any point.

In total, 160 students completed the onboarding survey, and 118 successfully configured the VPN on their primary devices. By the end of the three-week deployment, 121 participants submitted the exit survey, and 18 joined follow-up interviews. Participants ranged in age from 22–39 years ($M = 24.0$, $SD = 1.92$). The majority were graduate students in Computer Engineering ($n = 104$), with the remainder enrolled in Computer Science or Electrical Engineering. This study was approved by the Institutional Review Board (IRB).

\subsection{VPN-Based Monitoring System and Study}
\begin{figure}[t]
\centering
  \includegraphics[width=0.7\columnwidth]{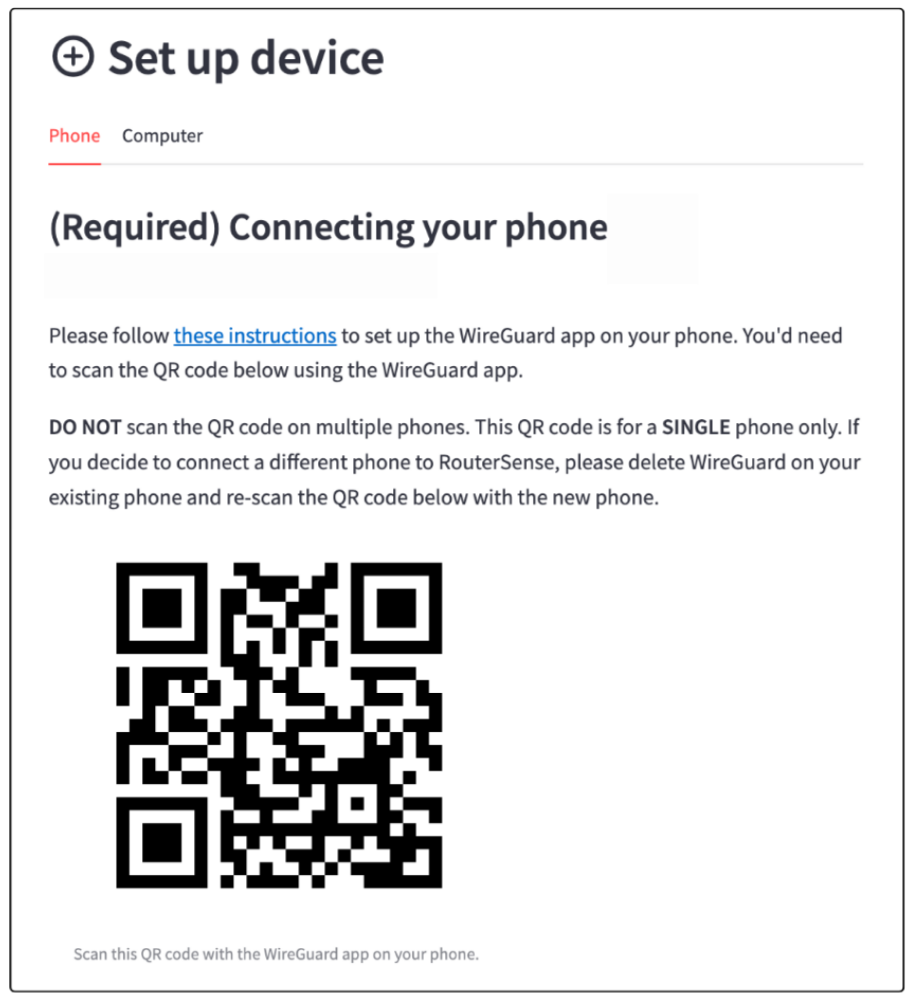}
  \caption{Participants install the official WireGuard client, scan a study-specific QR code generated by the portal, and the device is configured to route traffic through the research server.}
  \Description{}
  ~\label{fig:setup}
\end{figure}
\begin{figure}[t]
\centering
  \includegraphics[width=0.7\columnwidth]{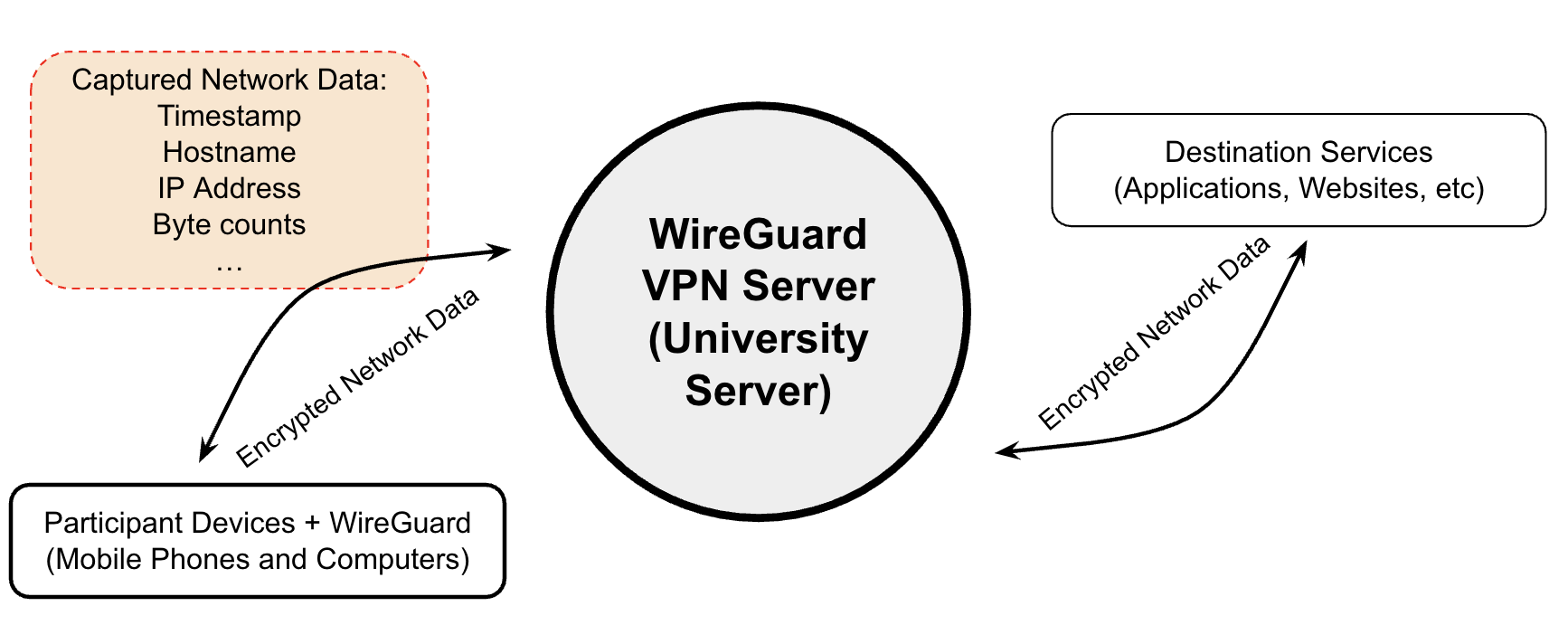}
  \caption{Schematic of the data path, illustrating that anonymous network traffic data is stored while application content remains unexamined.}
  \Description{}
  ~\label{fig:data_transfer}
\end{figure}

We instrumented participants’ devices using a VPN to record patterns of digital activity. For this study, we adopted WireGuard, a lightweight and widely used VPN protocol recognized for its stability, ease of setup, and compatibility across various operating systems. The enrollment process was designed to be straightforward. Participants downloaded the official WireGuard client from their app store and then scanned a study-specific QR code, shown in Figure~\ref{fig:setup}. This code, generated by our portal, automatically configured their device to connect through our research server. By relying on WireGuard’s standard application and eliminating the need for custom installations, we reduced both technical complexity for participants and potential concerns about researcher-built software. Once connected, the research server used the open-source tool \texttt{tshark} to record network metadata at frequent intervals.  Given the sensitivity of monitoring internet traffic, we incorporated safeguards at both the system and protocol levels. To prevent linkage between digital patterns and real-world identities, participants created their own identifiers (PIDs), which were never linked to personal information and could be regenerated at any time. The server retained only minimal metadata, including PID, IP address, hostname, timestamps, and traffic volumes, which were encrypted at rest on the institution's managed infrastructure, shown in Figure~\ref{fig:data_transfer}. No packet contents were collected, and access to metadata was restricted to the research team. For example, if a participant accessed ChatGPT, the log would record a connection to \texttt{chatgpt.com} and the associated data volume, but not the content of prompts or responses. Device IPs were used as unique identifiers for each enrolled device, enabling us to distinguish, for instance, between a participant’s phone and laptop. Importantly, these IP addresses did not correspond directly to individual users: a single participant could contribute multiple device IPs, and the addresses themselves were private (i.e., in the \texttt{10.0.0.0/8} private subnet), locally assigned values that cannot be used to de-anonymize participants. In practice, device IPs therefore functioned as device-level identifiers, allowing us to track usage across devices while maintaining minimal privacy risk.

We also emphasized transparency and participant control. Before enrolling, participants viewed a short orientation that included slides and a researcher-led discussion on risks and protections. During the deployment, they could disable the VPN whenever they wished, giving them direct authority over when logging occurred. By making participation opt-in and revocable, the study aligned with digital ethnographic principles that prioritize agency and autonomy. 

\begin{figure}[t]
\centering
  \includegraphics[width=\columnwidth]{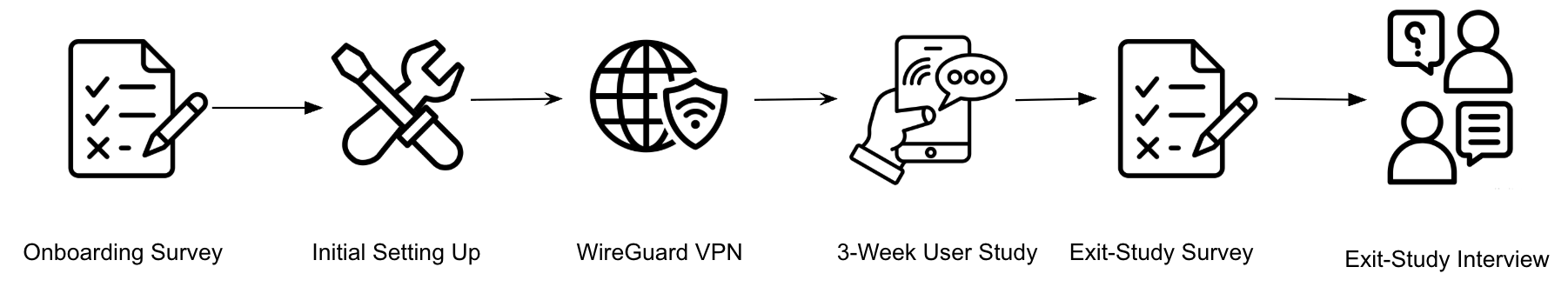}
  \caption{Overview of the three-week deployment with portal status checks, continuous VPN logging on participant devices, and post-study survey/interview.}
  \Description{}
  ~\label{fig:study_procedure}
\end{figure}

Following consent and onboarding, participants ran the VPN for a three-week deployment. The VPN was designed to remain active in the background, though reconnection was occasionally required after events such as device reboots. To reduce the chance of unintentional disengagement, the study portal sent periodic reminders to confirm VPN status. Figure~\ref{fig:study_procedure} provides a visual overview of the procedure. Compensation was provided in the form of course credit, allocated incrementally. Participants who met a minimum participation threshold—defined by sustained periods of active VPN use—received the full amount of credit. This structure balanced fairness with ongoing engagement, while avoiding overemphasis on payment as the primary motivator.


\subsection{Exit Measures}
At the conclusion of the three-week deployment, participants completed an exit survey and were invited to a follow-up interview. The survey incorporated two standardized instruments: the NASA Task Load Index (NASA-TLX)~\cite{hart1988development}, to capture perceived workload and burden, and the System Usability Scale (SUS)~\cite{brooke1996sus}, to assess the usability and “invisibility” of the VPN infrastructure. We also asked participants to identify the applications they considered most integral to their daily routines, providing self-reported context for interpreting logged activity.  

Semi-structured interviews were conducted over video calls to probe participants’ experiences in more depth. Questions addressed how they managed the VPN (e.g., whether it remained active or was occasionally disabled), how it fit into their everyday practices, and, if applicable, what motivated decisions to disconnect from it. We also explored their perceptions of the privacy safeguards and how these influenced trust and sustained engagement. These interviews were framed not only as an opportunity to capture participants’ practices and perspectives, but also as a way to reflect on the broader potential of VPN-based monitoring for studying sensitive domains of student life—particularly the everyday integration of AI into academic work and routines.

Finally, a subset of iPhone users opted to share Apple’s App Privacy Report (APR), which documents the domains contacted by installed apps over a seven-day window. These reports offered an additional lens for validating VPN logs, particularly for identifying AI-related services, while remaining entirely voluntary. We emphasized that APRs reveal only the domains associated with apps, not the content of interactions, and that declining to share them carried no impact on compensation or study participation.

\subsection{Data Analysis}

Building on the data sources described in Section 3, we analyzed VPN traffic logs, survey responses, and interview transcripts using complementary analytical approaches. The goal of this analysis was to transform raw network traffic traces, perception measures, and qualitative reflections into interpretable patterns of AI engagement and participant experience. Each dataset required a distinct analytic procedure, detailed below.

\subsubsection{Digital Trace Analysis}
VPN traffic was captured as individual network flows, defined as short-lived connections between a participant’s device and an online service. To study patterns of AI engagement, flows were aggregated into sessions using a five-minute inactivity threshold; sensitivity checks with 3- and 10-minute thresholds produced consistent results. Each session was then mapped to a specific AI tool domain. AI-related traffic was identified using a curated domain list derived from Apple’s App Privacy Reports (APRs), which document the most frequently contacted domains for installed applications. The final list included ChatGPT, Claude, Copilot, DeepSeek, Gemini, Grok, and Perplexity. No traffic corresponding to Grok was observed in our dataset, and it was therefore excluded from analysis. Gemini traffic, which appeared primarily through Google Search, was also excluded from aggregate usage totals to avoid inflating counts. All analyses were conducted at the device level (anonymized IP), since participants could register multiple devices (e.g., phone and laptop). Device identifiers were not linked to personal identities, ensuring that usage traces remained anonymous.

\subsubsection{Survey Analysis}
Survey responses were summarized using descriptive statistics. For workload, we reported means and distributions across the NASA-TLX subscales, which capture the perceived burden of continuous VPN monitoring. For usability, we calculated overall System Usability Scale (SUS) scores to assess how participants experienced the VPN system. These quantitative measures provide a baseline for interpreting participants’ lived experiences alongside qualitative findings.

\subsubsection{Interview Analysis}
All interviews were transcribed and analyzed using thematic analysis. Of the 121 participants, 18 completed an interview. We continued interviewing until reaching thematic sufficiency, where no substantially new insights emerged. Two authors coded transcripts independently, then met with the full research team to reach consensus on the final codebook. This process surfaced four overarching categories: usability and reliability, privacy and trust, behavioral adaptations, and motivations for continued use. These themes are presented in Sections 4.6 - 4.8.

\section{Results}
We structure the Results around our three research questions. RQ1 examines how network traffic analysis can serve as a lightweight ethnography; we address it by presenting our data processing pipeline and overall traffic patterns in Section 4.1. RQ2 focuses on usage patterns in students’ daily routines; we answer it through analyses of tool frequency and duration (Section 4.2), session behaviors (Section 4.3), and daily rhythms (Section 4.4). RQ3 concerns the lived experience of VPN-based logging; we answer it with survey and interview findings (Sections 4.5–4.9) that describe workload, usability, reliability, privacy, behavior changes, and willingness to continue use.

\subsection{Overall Network Traffic Volume Patterns by Participants}

Overall, across the three-week study, the VPN system logged 27.2 million individual network connections, or “flows.” A flow refers to a brief exchange of data between a device and an online service—for example, retrieving a webpage, refreshing an email inbox, or submitting a prompt to an AI tool. These connections came from 203 distinct devices (identified through anonymized IP addresses) and spanned more than 776,000 unique domains. 

To identify AI-related traffic, we leveraged Apple’s App Privacy Reports (APRs) provided by both participants and researchers. These reports document the domains that applications contact most frequently. Using APRs, we compiled a domain list corresponding to widely used AI tools—for example, \texttt{ws.chatgpt.com} and \texttt{chatgpt.com} for ChatGPT, and \texttt{suggest.perplexity.ai} and \texttt{perplexity.ai} for Perplexity. This process yielded a set of AI tools of interest: ChatGPT, Claude, Copilot, DeepSeek, Gemini, Grok, and Perplexity. Because our dataset contained no Grok-related traffic, we excluded it from further analysis.

Within the broader dataset, 1.32 million flows (4.9\%) were directed to AI services, including ChatGPT, Gemini, Claude, and Perplexity. Although most traffic involved non-AI services, AI tool usage represented a distinct and measurable share of participants’ overall activity. Of the 203 IP addresses observed, 180 (88.7\%) generated network traffic involving at least one AI service.

For analysis, we treated each distinct device IP address as a unique device. Participants could register up to two devices (e.g., a laptop and a phone), meaning that IP addresses map to devices rather than directly to individuals. To protect privacy, we did not maintain any records linking IP addresses back to specific participants; instead, de-identification ensured that analyses were conducted at the device level only.


\begin{figure}[t]
\centering
  \includegraphics[width=\columnwidth]{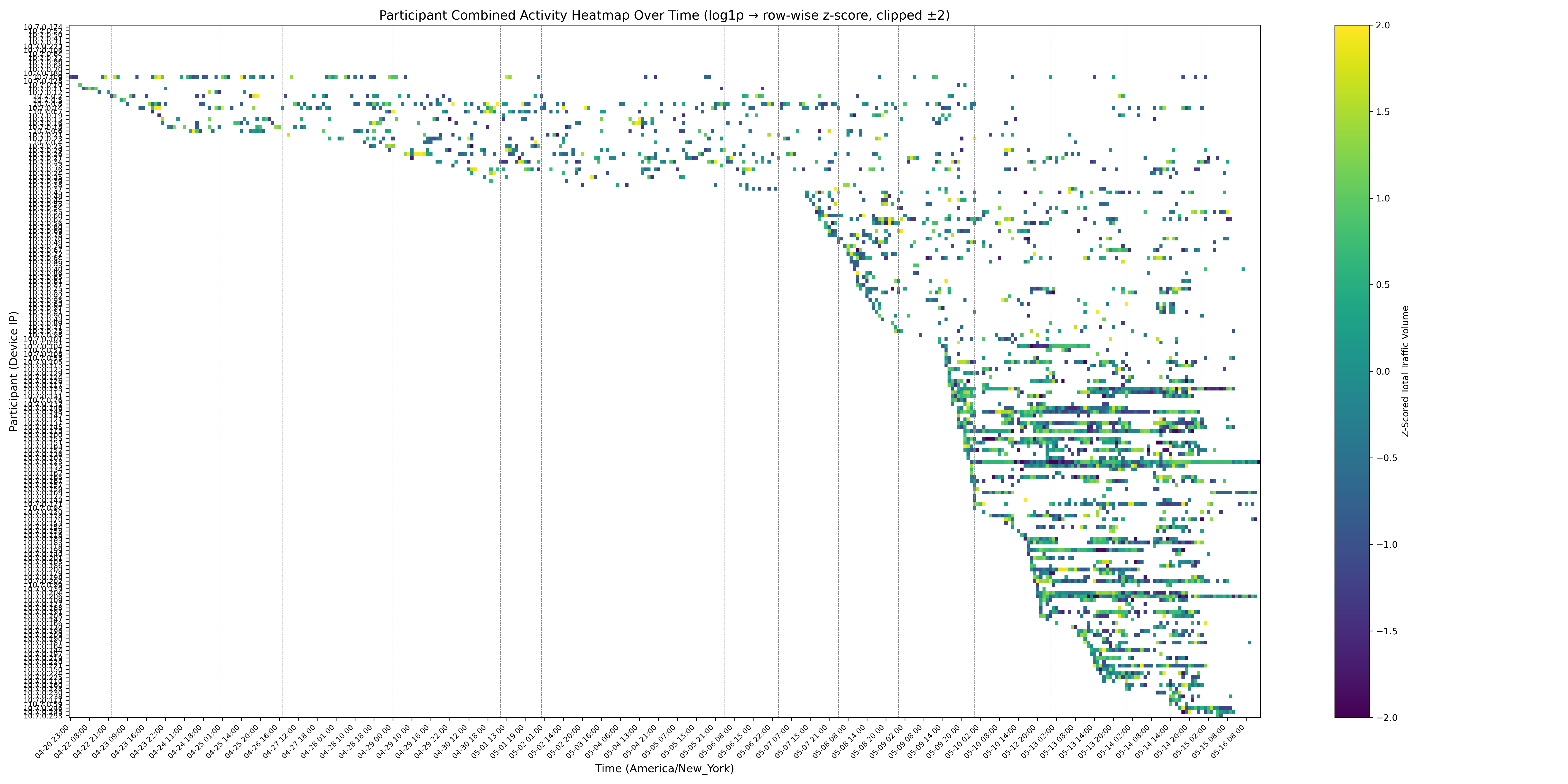}
\caption{Heatmap of participant-level AI activity over time. Each row corresponds to a single device, and each column to an hourly bin across the deployment. Color intensity reflects normalized traffic volume (row-wise $z$-scores), highlighting relative fluctuations while enabling comparison across participants.}
  \Description{}
  ~\label{fig:combined_heatmap}
\end{figure}

To visualize these patterns, we plotted each participant’s AI-related traffic volume over time (shown in Figure~\ref{fig:combined_heatmap}). Rows represent individual devices, and columns represent hours across the deployment. Traffic volumes were normalized per participant (row-wise $z$-scores), ensuring that relative fluctuations could be compared across participants without allowing heavy users to dominate the visualization. Brighter cells correspond to above-average activity for that participant, while darker cells reflect lower activity. 

The heatmap reveals several trends. Some participants generated steady streams of activity throughout the study, while others were more sporadic in their engagement. Drop-offs in specific rows suggest participants who disengaged or stopped reconnecting partway through the deployment. The staggered starting points visible across rows are a result of our deployment logistics. Due to server capacity limits, participants were onboarded in groups, which explains why some devices appear to have longer logging periods. Temporal clustering of bright regions also points to shared periods of higher activity, likely reflecting common study and work routines.

Together, these results highlight both temporal regularities and individual differences. While some participants relied heavily on AI tools as part of their daily practice, others engaged with them only minimally. This heterogeneity underscores how our method enables an overall understanding of AI usage at scale—capturing not only aggregate patterns but also the diverse ways students incorporate AI into their routines. In the next section, we narrow our focus to examine traffic involving AI services specifically.

\subsection{AI Tool Usage Frequency and Duration}

After identifying AI-related flows, we next examined usage at the device level to understand both how often and how long participants engaged with these tools. Specifically, we ranked devices by the number of AI services they accessed and highlighted the top 15 device IPs with the broadest engagement. This perspective captures not only the intensity of use but also the breadth of adoption, distinguishing participants who experimented with multiple tools from those who relied primarily on a single service. By focusing on frequency and duration, we can surface heterogeneous patterns of engagement—ranging from light, occasional use to heavy reliance—providing a more nuanced view of how AI tools were incorporated into daily routines.

\begin{table*}[t]
\centering
\scriptsize
\begin{tabular}{lrrrrrrrrr}
\toprule
\textbf{Device IP} & \textbf{ChatGPT} & \textbf{Claude} & \textbf{Copilot} & \textbf{DeepSeek} & \textbf{Gemini} & \textbf{Perplexity} & \textbf{Total AI-based Usage Count} & \textbf{Total AI-based Usage Time} \\
\midrule
10.7.0.123 & 15371 & 0 & 3 & 0 & 80531 & 0 & 15383 & 1d 18h \\
10.7.0.204 & 1839 & 0 & 699 & 0 & 54656 & 0 & 2538 & 0d 7h \\
10.7.0.209 & 780 & 0 & 82 & 134 & 52370 & 0 & 996 & 0d 2h \\
10.7.0.183 & 0 & 0 & 595 & 0 & 48154 & 0 & 595 & 0d 1h \\
10.7.0.146 & 915 & 294 & 252 & 183 & 47064 & 0 & 1644 & 0d 4h \\
10.7.0.126 & 1816 & 0 & 433 & 0 & 43186 & 0 & 2249 & 0d 6h \\
10.7.0.199 & 4 & 0 & 4538 & 98 & 38036 & 0 & 4687 & 0d 13h \\
10.7.0.206 & 6439 & 0 & 24 & 37 & 35429 & 0 & 6507 & 0d 18h \\
10.7.0.112 & 1359 & 0 & 160 & 0 & 31314 & 0 & 1519 & 0d 4h \\
10.7.0.148 & 3575 & 0 & 671 & 16 & 27977 & 0 & 4262 & 0d 11h \\
10.7.0.152 & 1401 & 0 & 655 & 522 & 29291 & 0 & 2578 & 0d 7h \\
10.7.0.104 & 5043 & 0 & 54 & 0 & 24519 & 0 & 5097 & 0d 14h \\
10.7.0.133 & 801 & 216 & 262 & 0 & 25546 & 61 & 1340 & 0d 3h \\
10.7.0.191 & 473 & 0 & 74 & 15 & 25518 & 0 & 562 & 0d 1h \\
10.7.0.121 & 337 & 0 & 100 & 0 & 24129 & 0 & 437 & 0d 1h \\
\bottomrule
\end{tabular}
\caption{Per-tool usage counts and aggregate AI usage for the top 15 device IP addresses. 
\textbf{Note:} AI totals exclude Gemini to avoid double-counting with Google Search, which may already integrate Gemini in the backend.}
\label{tab:combined_ai_usage}
\end{table*}

As we have 203 distinct devices in the dataset, we selected the top 15 IP addresses for closer examination. Table~\ref{tab:combined_ai_usage} presents their per-tool usage, revealing several distinct patterns. Although we display example IP addresses in tables, these are private local 
addresses (e.g., 10.x.x.x) rather than public routable ones. Such addresses 
function only as internal identifiers within the VPN and cannot be used to 
trace participants on the open internet, effectively serving as device IDs. 
This minimizes any privacy risks associated with presenting IPs in the paper. The data shows a wide spectrum of engagement, even among these top users. Some devices generated exceptionally high volumes of AI traffic—one exceeding 80,000 requests and accumulating more than 11 days of connection time, while others demonstrated more moderate activity. We also observed a significant variation in the distribution of use across different AI tools. While participants accessed ChatGPT from many devices, its volume remained moderate. In contrast, Gemini traffic overwhelmingly dominated the activity for nearly all of these heavy users.

This imbalance arises from a key methodological decision. We noticed that Google has integrated Gemini-powered summaries directly into its search engine results. Consequently, we included traffic to \texttt{google.com} as evidence of AI interaction, which significantly inflated Gemini’s measured presence compared to standalone tools like Claude, Copilot, or Perplexity that require users to navigate to their applications or websites explicitly.

These findings offer two essential takeaways. First, analyzing network traffic allows us to capture both the breadth of AI adoption, including the number of different tools people access, and the depth of their engagement, including the intensity and duration of use over time. Second, the Gemini case highlights how generative AI is increasingly blending into everyday digital routines. Participants often encountered AI not by deliberately seeking out a specific tool, but as an embedded, infrastructural feature of a familiar platform. This distinction is important: when AI becomes infrastructural rather than optional, researchers must interpret "use" differently, and designers must consider how such invisible integrations shape user perceptions of agency and engagement.

\subsection{Session-Based Patterns of AI Tool Use}
To understand the temporal dynamics of engagement, we moved beyond raw traffic counts to aggregate individual network flows into discrete user sessions. In this context, a flow represents a single, continuous network connection between a participant's device and an AI service, typically corresponding to one request-and-response cycle. We define a session as a constant period of interaction with a single AI tool, demarcated by at least 5 minutes of inactivity. Because there is no universally established inactivity threshold for defining sessions in AI tool usage, we experimented with several candidates (3, 5, and 10 minutes). We chose 5 minutes as our primary definition, as it provides a balance between capturing brief, micro-interactions and avoiding fragmentation of continuous activity. This choice aligns with common practice in sessionization studies of web and app usage logs, while sensitivity checks with 3- and 10-minute thresholds yielded consistent patterns. This framing allows us to analyze AI use as distinct episodes of engagement. For example, if a participant sent three queries to ChatGPT at 2:00, 2:01, and 2:03 PM, we treated these as part of the same session, which lasted 4 minutes. If the following query came at 2:25 PM, more than 5 minutes later, we counted that as the start of a new session.

\begin{figure}[t]
\centering
  \includegraphics[width=0.8\columnwidth]{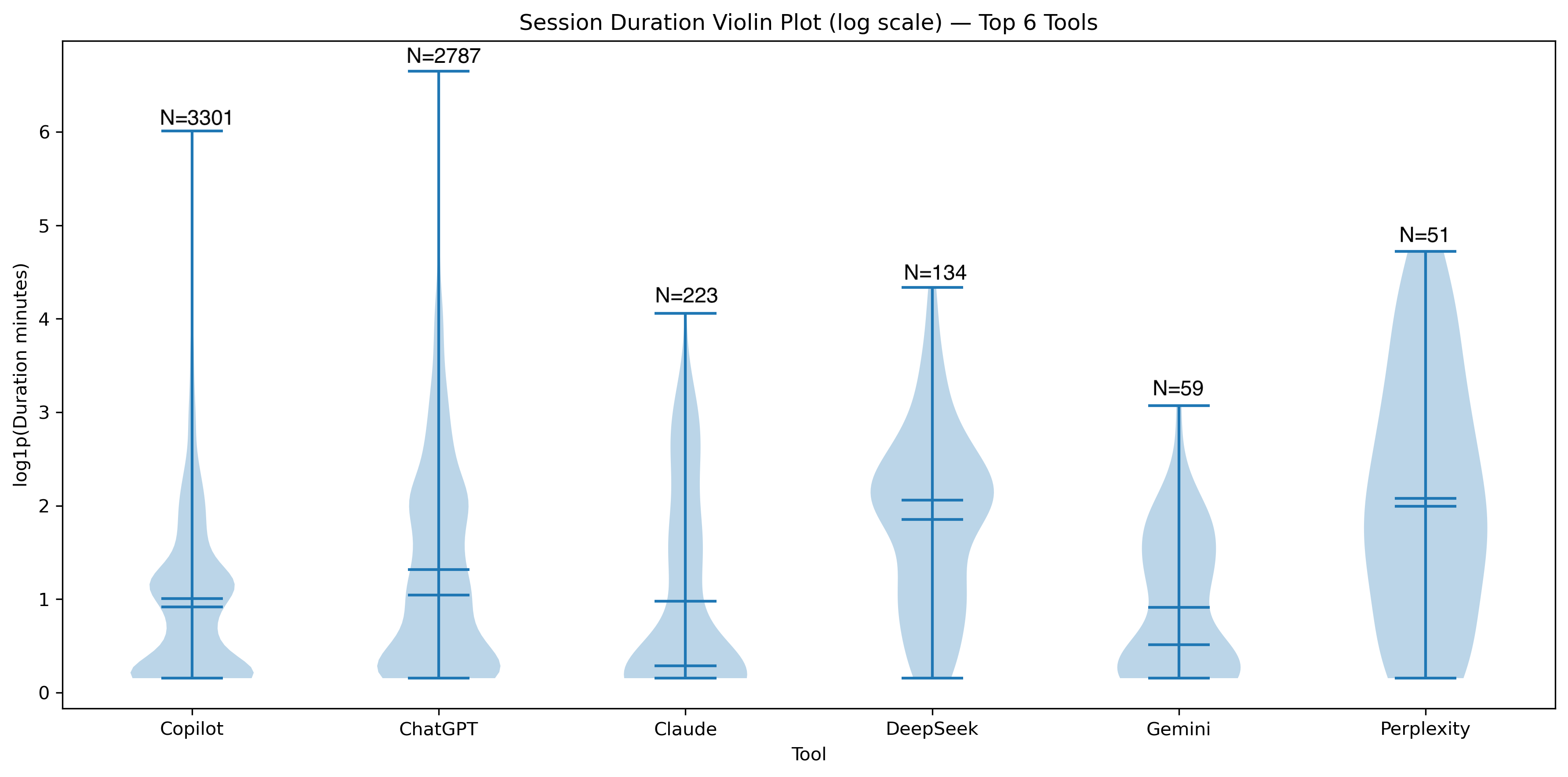}
  \caption{Distribution of session durations in minutes for six generative AI tools. Each histogram shows a long-tail distribution, with a high concentration of sessions lasting less than 10 minutes.}
  \Description{}
  ~\label{fig:viloin_chart_session}
\end{figure}

Figure~\ref{fig:viloin_chart_session} visualizes the distribution of these session lengths across the most popular tools. The data reveals a clear and consistent pattern: the vast majority of AI sessions were brief, lasting only a few minutes, followed by a steep drop-off in frequency for longer durations. This short-burst interaction model holds across all tools, from dedicated platforms like ChatGPT and Copilot to the less frequently used Claude and Perplexity. Notably, even Gemini, which dominated overall traffic due to its integration with Google Search, exhibits the same distribution of many short sessions and very few extended ones.

These findings strongly suggest that participants integrate AI tools into their daily routines as lightweight, on-demand resources rather than engaging in long, sustained blocks of work. A typical interaction appears to be a quick query—to clarify a concept, generate a code snippet, or check a summary—before the user returns to another primary task. From a digital ethnography perspective, this session-based analysis illuminates how AI tools fit into the natural rhythms of academic work: they function not as destinations for prolonged activity, but as embedded cognitive aids accessed in rapid, intermittent bursts throughout the day.


\subsection{Daily Rhythms of AI Tool Engagement}

\begin{figure}[t]
  \centering
  \includegraphics[width=0.9\linewidth]{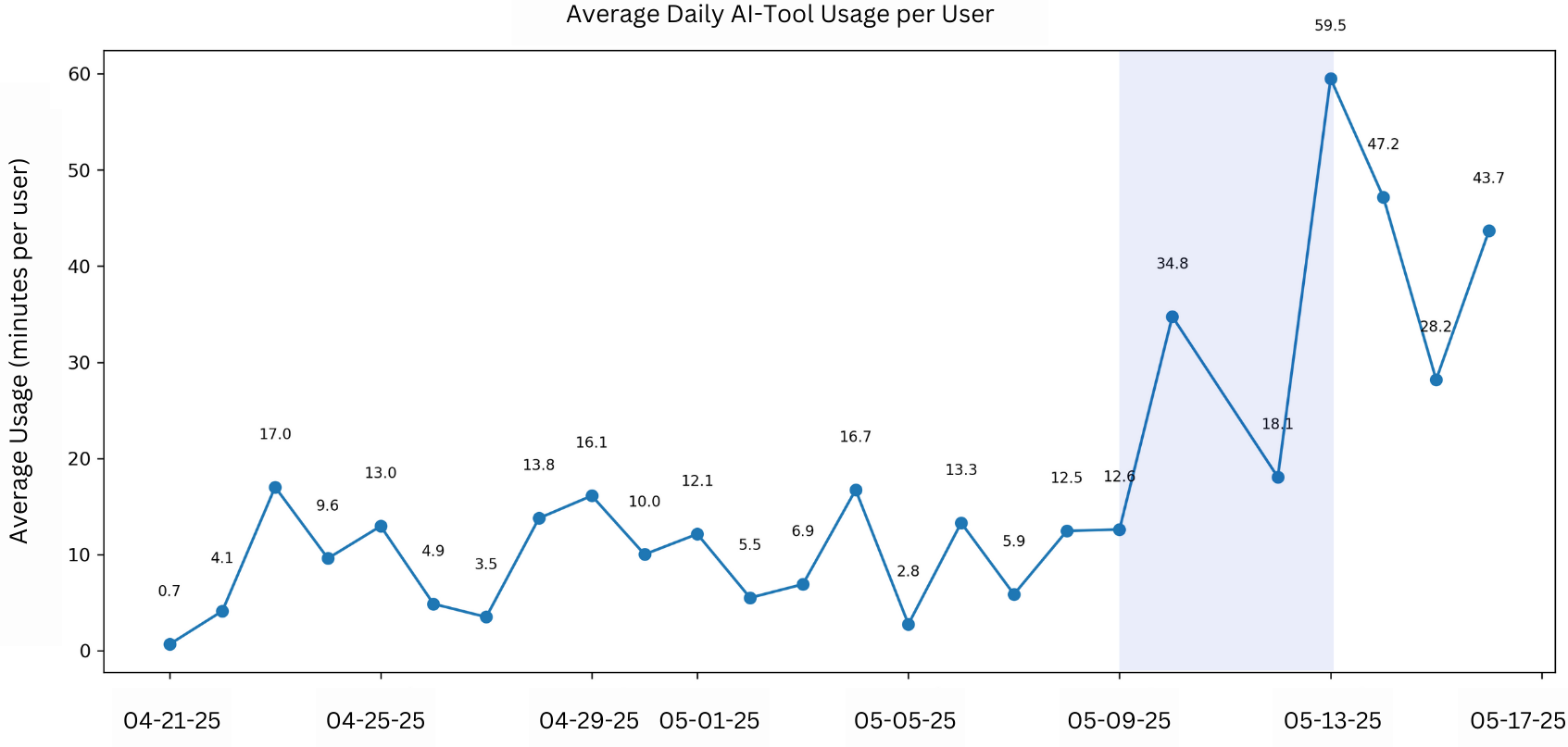}
  \caption{Average daily AI-tool usage per participant device across the study period. We compute the daily average by aggregating AI-related activity for each distinct device (identified by anonymized IP), then averaging across all devices active on that date. “AI-related activity” reflects requests to a curated set of AI domains.}
  \label{fig:line_chart_label}
\end{figure}

Participants did not all begin using the VPN on the same day, so absolute traffic counts do not fully capture collective behavior. To address this, we calculated the average daily usage per participant, which reveals clearer temporal patterns. As Figure~\ref{fig:line_chart_label} shows, AI tool activity increased sharply from May 9 to 14, coinciding with the last week of the semester. And May 14th was the last day of the final class exam. Students engaged with AI tools more heavily during this period, with average daily usage rising from roughly one hour to over seven hours per participant at the peak. This surge suggests that students relied on AI tools as on-demand support during periods of heightened academic pressure, such as exam preparation and final project deadlines.

\begin{figure}[t]
  \centering
  \includegraphics[width=0.95\linewidth]{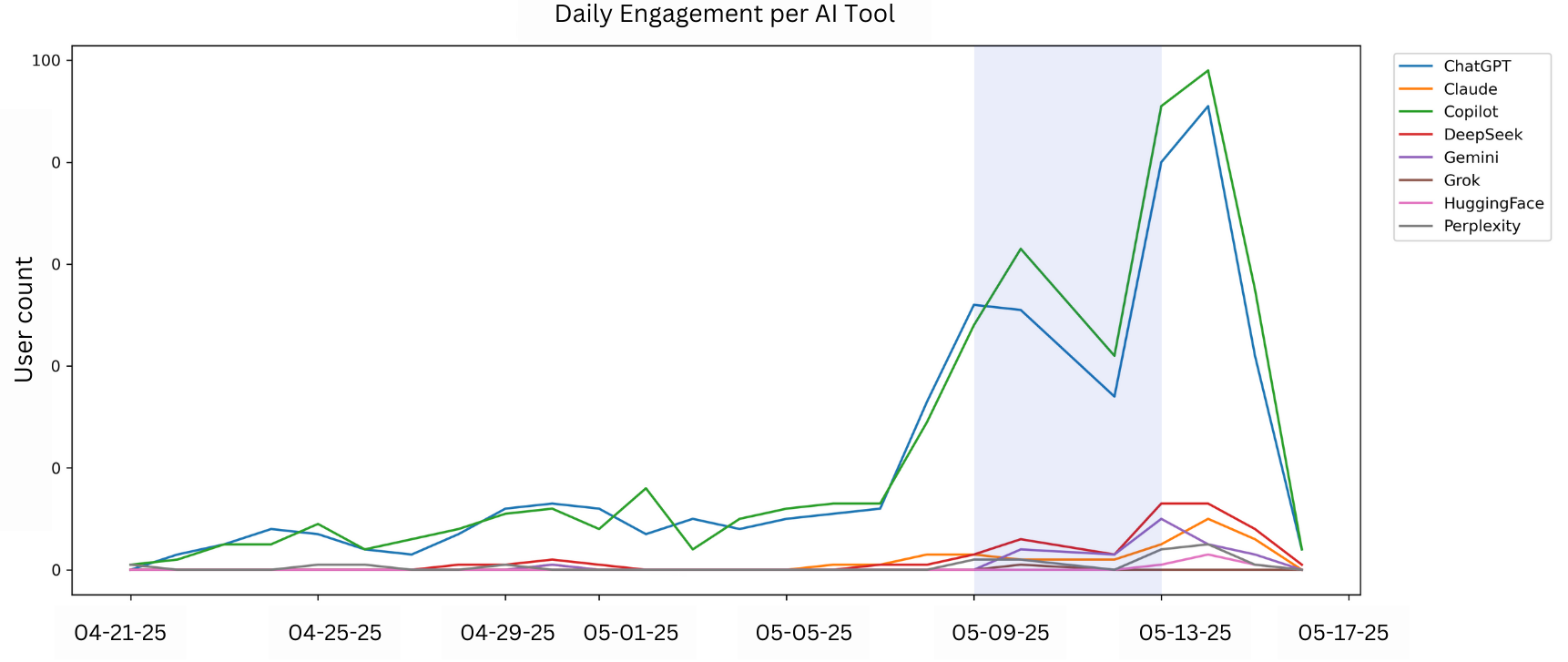}
  \caption{Daily count of unique participant devices exhibiting AI-related activity. We treat each anonymized IP as one device. A device is counted as “active” on a day if it generates at least one AI-related request to the curated domain list. This series shows the breadth of participation by day, complementary to the per-device averages in Fig.~\ref{fig:line_chart_label}.}
  \label{fig:line_chart_overall}
\end{figure}

Daily counts of unique devices echo this trend (Figure~\ref{fig:line_chart_overall}). More students turned to AI tools during the same period, producing a simultaneous rise in both intensity (minutes per user) and breadth (number of active users). Usage fluctuated earlier in the study but became more concentrated as academic demands increased. Taken together, these patterns highlight the role of AI tools as flexible, situational resources that students draw on most when workloads spike, rather than as tools of steady, everyday reliance.

Interestingly, while the network logs clearly show a sharp increase in AI tool activity during the final week of the semester, few participants explicitly reported altering their usage during this period. This gap between behavioral traces and self-reflections highlights a key limitation of self-report methods: students may not fully recognize, recall, or disclose changes in their own practices. Our VPN-based logging approach, by contrast, surfaces these shifts directly through digital traces. We revisit this discrepancy in Section~4.8, where we examine participants’ reflections on monitoring and the ways they understood (or overlooked) their own behavioral adaptations.

\begin{figure*}[t]
  \centering
  \includegraphics[width=0.8\columnwidth]{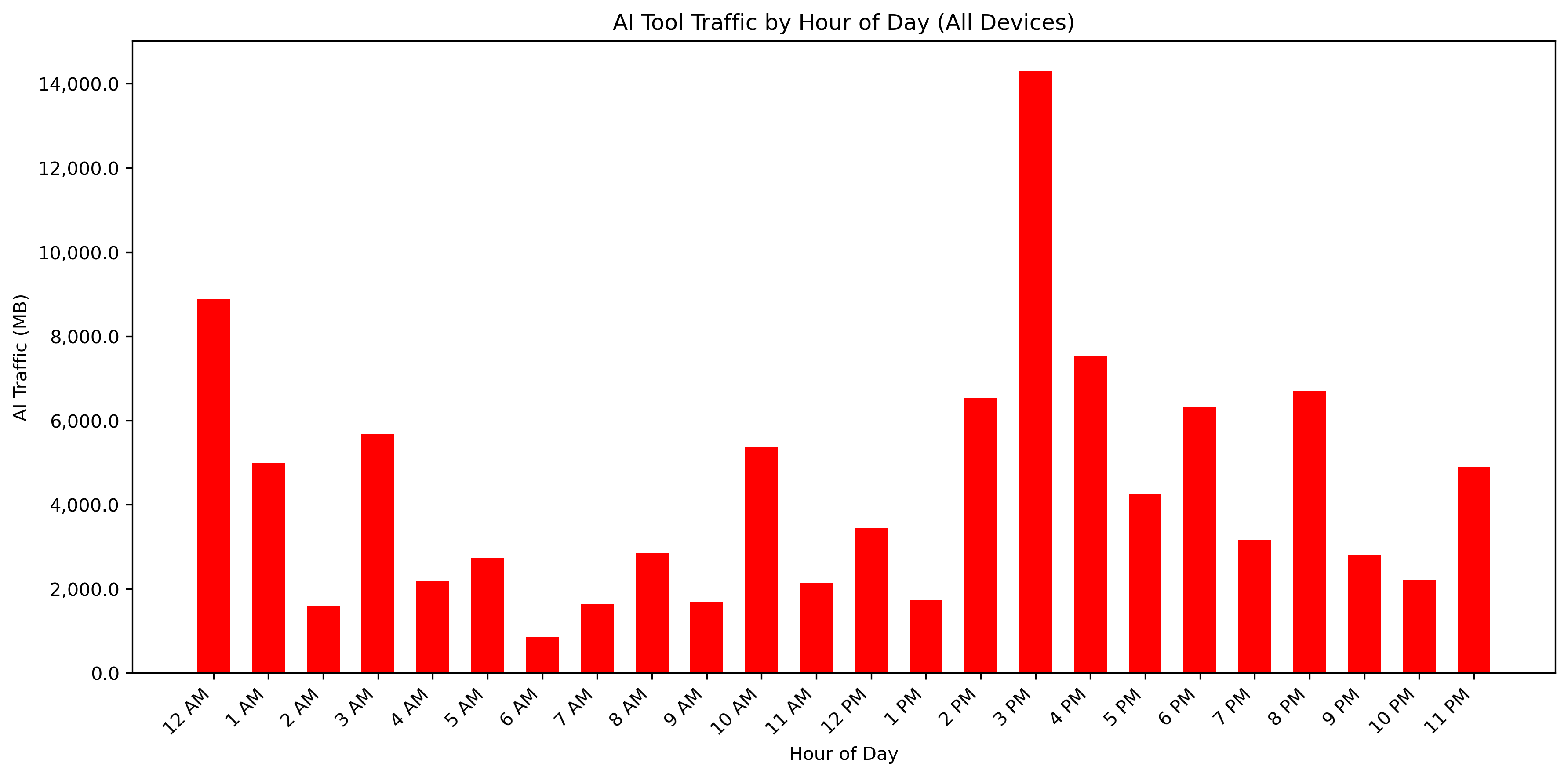}
  \caption{AI-related network traffic by hour of day across all study devices (times shown in Eastern Daylight Time, EDT). Bars show the total volume of AI traffic (MB) aggregated over the deployment for each local hour (0–23). We compute volumes from requests to the curated AI domain list and sum upstream+downstream bytes per hour.}
  \label{fig:ai_hour_usage}
\end{figure*}

In addition, we aggregate AI-related traffic across all devices and categorize it by local hour of day to reveal daily usage patterns (Figure~\ref{fig:ai_hour_usage}). Students engage with AI tools around the clock, with clear peaks in the afternoon and evening. Traffic rises through the day, spikes in the mid-afternoon (3–4 PM), and remains high into the evening (6–9 PM). We also see a smaller late-night shoulder (around 1 AM), while early-morning hours show the lowest volumes.

These patterns show how students fold AI into everyday work: they turn to tools during busy afternoon blocks, continue into the evening, and still generate activity late at night. Passive VPN logging lets us recover these around-the-clock rhythms without asking students to repeatedly recall or self-report brief, distributed interactions—something traditional surveys or diaries struggle to capture at this granularity and cadence. This analysis complements digital ethnography by surfacing frequent, in-situ activity that would otherwise go unreported.

\begin{figure}[t]
  \centering
  \includegraphics[width=0.7\columnwidth]{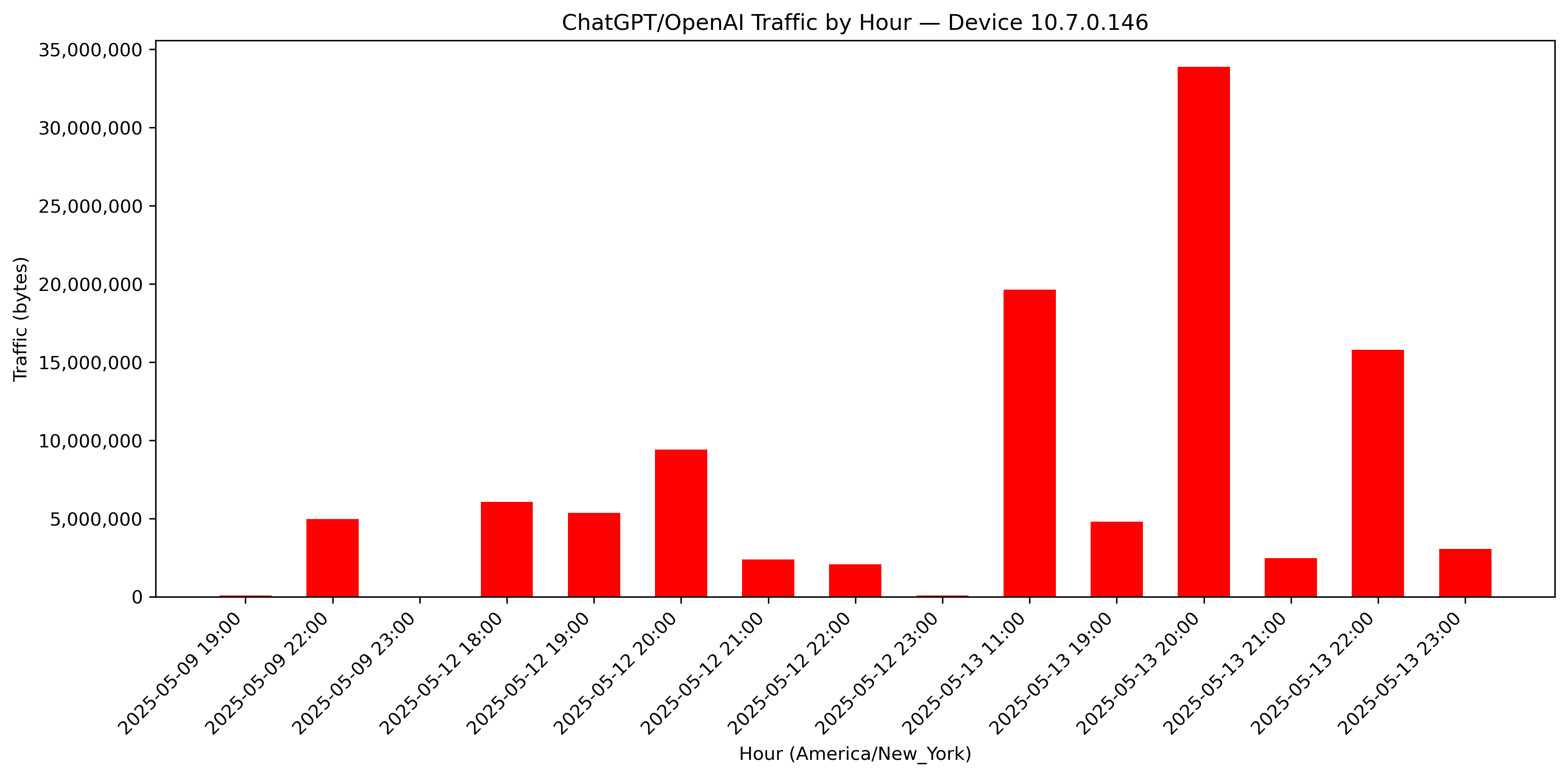}
  \caption{Hourly traffic volume to OpenAI/ChatGPT domains for a single device (anonymized IP 10.7.0.146) during active study days. Bars show the sum of upstream+downstream bytes per local hour, computed from requests to the curated ChatGPT domains.}
  \label{fig:device_chatgpt_hourly}
\end{figure}

To ground the aggregate trends in an individual trace, we examine one device (anonymized IP 10.7.0.146) and plot its ChatGPT/OpenAI traffic by hour (Figure~\ref{fig:device_chatgpt_hourly}). This device shows concentrated bursts in the afternoon and evening, including a large block of traffic around 3–4 PM and 8–10 PM on May 13, with lighter activity late at night and minimal activity early in the morning. This pattern mirrors the cohort-level results: students fold AI use into late-day work cycles and ramp up activity during the final week of the term. The device-level view also illustrates the value of VPN-based sensing for digital ethnography. We can recover within-day rhythms and short, distributed episodes of use without requiring participants to repeatedly recall or log micro-interactions—a task that diary and survey methods struggle to capture at this level of granularity.

\subsection{Post-Study Survey: Workload and Usability Ratings}
After completing the study, participants filled out a post-study survey that included the NASA Task Load Index (NASA-TLX) and the System Usability Scale (SUS) for the WireGuard-based VPN system. These measures contextualize the user experience in terms of perceived effort and usability.

\begin{figure}[t!]
\centering
  \includegraphics[width=0.5\columnwidth]{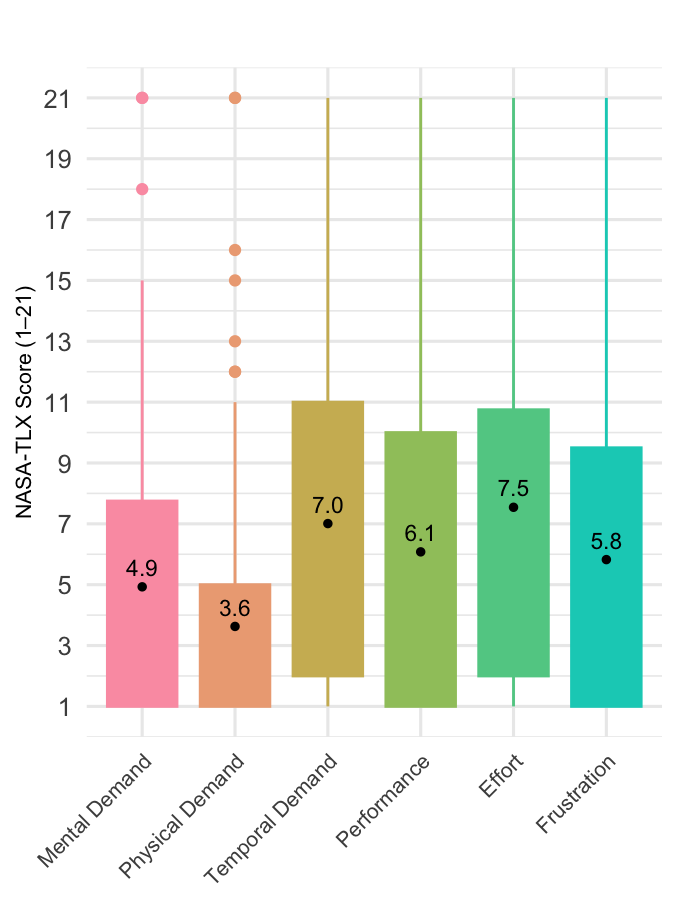}
  \caption{NASA-TLX workload ratings across six dimensions (1–21 scale, higher scores indicate greater perceived demand except for performance).}
  \Description{}
  ~\label{fig:box_plot}
\end{figure}

On the NASA-TLX scales, participants reported generally low workloads. Mental Demand averaged 4.93 ($SD=4.41$), Physical Demand 3.63 ($SD=4.18$), and Temporal Demand 7.01 ($SD=6.14$). Performance ratings averaged 6.08 ($SD=6.51$), while Effort was 7.54 ($SD=5.81$) and Frustration was 5.82 ($SD=5.77$). Figure~\ref{fig:box_plot} visualizes these distributions. The chart shows that Physical Demand was consistently rated as low across participants, while Temporal Demand, Effort, and Performance were slightly higher but remained moderate. Frustration scores were clustered toward the low end, indicating that the system did not create notably high levels of annoyance or disruption. The elevated Effort and Frustration of an unexpected server outage can explain ratings during the deployment, which caused some participants to check the VPN status more frequently and worry about the stability of the data connection.

Participants’ usability judgments reinforced these findings. The mean SUS score was 76.2 ($SD=15.6$), well above the benchmark score of 68, which indicates “above-average” usability. Individual scores ranged from 30 to 100, but most participants rated the system between 70 and 85, corresponding to “good” usability on the SUS adjective scale. This result suggests that most participants found the WireGuard-based VPN system to be a highly usable and acceptable tool for tracking their network activity. Qualitative comments supported this perception: participants emphasized the simplicity of installation and ease of use, with one noting,  “\textit{The entire interface and everything was incredibly simplified},”. Only a few participants reported minor disruptions when the VPN server temporarily went offline. Yet, these did not significantly affect their overall usability ratings, “\textit{Once or twice when the VPN server went down, it just interfered with my network}”.  More detailed quotes and usability findings are presented in Section~\ref{subsec_usability}. Taken together, the post-study survey data indicate a positive user experience: participants reported low workload, regarded the system as easy to use, and remained highly engaged throughout the study. This combination of low perceived effort and strong usability likely contributed to the high completion rates observed in the deployment.

\subsection{Usability and System Reliability}\label{subsec_usability}
Most participants (15 of 18) reported that the VPN-based monitoring system was easy to use. They found the installation and daily operation to be straightforward, with minimal disruption to normal device use and network connections. Participants highlighted the intuitive user experience on both mobile and desktop platforms. As one explained:

\begin{quote}
[P1]``The system is easy to use. It’s easy to follow the directions… on phone and computer.''
\end{quote} 
Another participant echoed this sentiment:
\begin{quote} 
[P8]``The entire interface and everything was incredibly simplified, and I did not have any issues with the system itself.''
\end{quote}

Despite the overall positive usability comments, minor technical problems were noted. A few participants experienced intermittent connectivity issues when the system's VPN or server was temporarily unavailable. For instance, one participant (P2) described that \textit{"once or twice when the [system] server went down, it just interfered with my network. That was the only problem I faced.''} In some cases, switching between Wi-Fi and cellular networks caused brief connection failures. Notably, most participants chose not to disable the VPN when such issues occurred, either because the problems were infrequent or because they were motivated to maintain data collection. One participant (P13) explained that even when connections failed, they \textit{``never turned it off… I just thought of other ways to get rid of it.''}, rather than disconnecting.

This finding suggests that the system’s “set-and-forget” design minimized cognitive overhead, which created tolerance for minor disruptions. Unlike diary studies or experience sampling, which demand repeated user input, this VPN approach shifted the effort to a single upfront installation and then operated unobtrusively in the background. 
As a supplement to traditional ethnographic methods, such an approach is valuable precisely because it captures traces of participants’ frequent, everyday activities with minimal burden. Also, this demonstrates that the strength of VPN-based monitoring lies in lowering ongoing demands to observe authentic daily practices at scale.

\subsection{Privacy Concerns and Trust in Data Handling}
Privacy emerged as a significant theme, with 12 of 18 participants expressing initial unease about continuous monitoring. Several participants described a feeling of being "watched" by the system, which made them cautious. For example, one participant (P2) admitted to having the feeling \textit{``it is just like, in the back of the mind… someone is monitoring my activities.''} 
This sense of surveillance prompted worries that sensitive information, such as banking details, might be inadvertently recorded. As participants explained, 
\begin{quote}
[P5]``I will be a little bit worried when I enter some personal information… maybe I would want to turn [the VPN] off for those cases.''
\end{quote}

The level of privacy concern often correlated with the participant's trust in the research context. Many took comfort in knowing the study was for academic purposes. For instance, this user stated:
\begin{quote}
    [P6]``the data you record… is for academic purpose, so it’s okay for me. But if others let me use this… in a commercial background, I will not trust this.''
\end{quote}

This sentiment was common; participants granted trust due to the university affiliation and consent procedures, but would be far more wary of a similar but commercialized application. Transparency at the start of the study, such as explaining that only metadata would be collected, helped alleviate concerns, including clarifying that only metadata, not content, was being collected. 
 One participant (P1) noted that after reading the study description, \textit{``I really didn’t care about any [privacy issues] because they knew the WireGuard[VPN] would not leak personal data or link data back to them''.}
By the end of the study, several participants reported their initial worries had diminished or “gradually disappeared” as they became accustomed to the tool.

Trust in the system stemmed not only from technical safeguards but also from the broader research context. Participants maintained the VPN connection because they trusted the academic framing, understood that only metadata was collected, and knew they retained the option to disconnect. Initial anxieties faded as they became accustomed to the tool, though concerns about sensitive activities lingered. 
As a complement to digital ethnography, VPN-based methods benefit from their ability to record everyday digital traces, but their acceptance depends on cultivating trust through transparency and user agency. 

\subsection{Behavior Changes and Self-Reflection Under Observation}
Awareness of being monitored led over half of the participants (10 of 18) to alter their online behavior, particularly during the early days of the study. The most common reaction was self-censorship. Participants avoided websites or activities they perceived as too personal or embarrassing. For instance, one participant explained, 
\begin{quote}
[P9] ``I might consider I won’t visit some webpages I might not want people to see… I just will behave a little bit better when the tracker is on.''
\end{quote}
Another (P4) reported that, at the experiment’s outset, \textit{``when doing some web searching, I would use an incognito mode – which I never did before – to try to be more private.''} In a similar vein, P9 admitted deliberately avoiding adult content: \textit{``if I decide to… browse some weird website – let’s be honest, maybe a porn website – I would turn it off.''} Together, these accounts highlight a heightened self-regulation: participants modified their online practices, either by abstaining from certain activities or by adopting additional privacy measures, in direct response to the awareness that their network data was under observation.

Beyond changes in browsing habits, some participants developed new routines to ensure compliance with the study. Several became vigilant about maintaining the VPN connection, incorporating device checks into daily life. As one noted:
\begin{quote}
[P13] ``I always look at the status bar on my phone to see if the VPN icon is on. At home, I would check the WireGuard [VPN] regularly, like every two hours, to see whether the cumulative time was increasing.''
\end{quote}
Similarly, P9 remarked, \textit{``every day I woke up, the first thing was to check if WireGuard is on… and the last thing I do before bed is also to check if the VPN is on.''} Such practices illustrate how study incentives produced novel behaviors tied to the infrastructure itself—for example, leaving devices powered on longer than usual to maximize data collection time.

This effect, however, was not universal. For some, heightened caution faded with time, leading to habituation. P4 reflected: \textit{``I just realized I forgot all about this [being monitored]. I just used my device as usual.''} Others described more enduring adaptations, such as routine VPN checks or modified browsing. Still others expressed comfort with persistent monitoring, particularly when they perceived personal benefit. P2 stated he would be willing to keep the VPN running “forever” if it offered a 24-hour summary of background app activity. P12 likewise reported no discomfort: \textit{``I don’t think I need to turn it off at any time… It doesn’t affect me that much.''} These differences underscore the heterogeneity of privacy perception: while many practiced self-censorship or routine adaptations, a subset normalized or even embraced monitoring when they deemed the captured data non-sensitive or already public.

Overall, participant feedback shows that surveillance awareness produced tangible behavioral shifts, at least in the short term. These patterns resonate with the Hawthorne effect: awareness of observation initially disrupted routines, though many later reported reverting to baseline~\cite{adair1984hawthorne}. From an ethnographic standpoint, this illustrates how infrastructures of visibility restructure daily practices—prompting temporary reconfigurations that may give way to normalization. For HCI, the implication is clear: logging methods must anticipate cycles of adaptation and habituation, both of which shape user experience and the behavioral data collected.

At the same time, we hypothesize that our system \textit{presumably} reduces the Hawthorne effect relative to more obtrusive methods, because its passive, infrastructure-based design minimizes the cognitive overhead associated with observation. We stress that this remains an assumption rather than an empirical finding; confirming or falsifying it will require future studies with larger and more diverse cohorts. Our contribution here is therefore exploratory: this study represents a first step, providing initial evidence on the feasibility and acceptability of VPN-based behavioral monitoring.

We note three caveats in interpreting these findings. First, only 18 of 121 participants completed exit interviews, raising the possibility of selection bias. Second, the sample was drawn from a cybersecurity course, where students likely have heightened privacy awareness compared to peers, potentially skewing attitudes. Third, participants were international students from diverse countries, and prior work demonstrates that privacy attitudes vary significantly across cultures \cite{vitak2023data}, suggesting limited generalizability.  

Taken together, these findings demonstrate that passive, infrastructure-based monitoring influences behavior in diverse ways—from short-term self-censorship to normalization and even enthusiastic acceptance when framed as applicable. While our evidence is drawn from a specific, privacy-primed, and culturally diverse cohort, it illustrates the feasibility of VPN logging as a scalable, minimally intrusive extension of ethnographic methods, laying groundwork for future studies that can more systematically evaluate its impact on reactivity and long-term acceptability.

\section{Discussion}

\subsection{The Need for Passive and Unobtrusive Approaches}

Our findings reinforce the importance of passive, unobtrusive data collection methods in studying everyday technology use. Reliance on self-reports alone is problematic because frequent, routine digital interactions often escape accurate recall, and responses can be skewed by social desirability biases~\cite{gold2015validity, al2023designing}. This is especially true as AI tools become seamlessly woven into daily life; students may have numerous brief encounters with AI assistants throughout the day that they hardly notice, let alone remember to log. Capturing such micro-interactions requires methods that operate in the background without burdening the user.
Various forms of digital metadata offer this capability, including device logs, app usage trackers, keystroke records, and periodic screenshots of activity~\cite{onnela2016harnessing, hu2024exploring}. In the context of AI tools, a student might downplay using an assistant for homework due to fear of judgment, or simply forget dozens of minor tool uses that occurred outside of class.

Beyond accuracy issues, traditional methods are often obtrusive and burdensome when trying to capture continuous daily activities. Diary studies and experience sampling require frequent user input, which can disrupt natural behavior and lead to participant fatigue. Research in personal informatics has documented that intensive self-tracking quickly becomes a burden; users tire of constant journaling and many abandon manual tracking ~\cite{zhang2016examining, swain2023leveraging}. By minimizing reliance on human memory or compliance, passive methods can capture a more naturalistic record of behavior. This is where network traffic analysis offers a crucial advantage: it logs tool use automatically in the background, avoiding interrupting users and bypassing self-report biases. By integrating such passive measures, researchers can obtain a ground-truth record of behavior that complements or even replaces self-reported data. It can also produce rich behavioral traces that overcome the biases of self-report, yet remain lightweight in deployment. 

\subsection{Network Traffic as a Window into Daily Routines}
While our study focused on AI tool engagement, the underlying approach, analyzing network traffic, opens a window onto broader aspects of students’ daily routines. Internet usage patterns tend to mirror real-world behaviors, which means network data can reveal traces of activities beyond the specific apps or domains of interest. For instance, the flow of network connectivity over 24 hours can be a proxy for sleep and wake cycles~\cite{jaisinghani2023packets, mahmood2024routersense, abdullah2014towards}. If a student’s devices show virtually no outgoing traffic during certain late-night hours, it strongly suggests they were offline and likely asleep. In our context, this implies that AI usage data could be correlated with rest-activity rhythms. E.g., heavy AI tool use late at night might indicate last-minute study sessions, potentially flagging irregular sleep habits or heightened academic stress. More generally, we foresee that if we notice consistent periods of digital downtime (or conversely, intensive usage) that align with meal times, exercise routines, or social activities, it suggests that digital traces bridge domains of life. Our network-based approach thus has the advantage of capturing behavior in situ across different contexts. It paints a timeline of the user’s day: from academic work with AI platforms to entertainment, socializing, and rest. These insights underscore a key point: techniques like ours need not be confined to a single behavioral category. The same passive data streams can simultaneously inform multiple facets of user life.

\subsection{Activities Across Platforms and Devices}
This study shows that network traffic data emerges as a powerful and scalable resource for studying technology use across devices and applications. Unlike data from a single app or self-report on a single device, network logs can bridge across platforms, giving a unified view of user behavior in a heterogeneous tool ecosystem~\cite{wilson2021cross}. Our work highlights that focusing on a single platform can miss the broader picture. For example, if a student alternates between a laptop and smartphone (or between different AI applications), traditional instrumentation would require multiple tracking mechanisms.
In contrast, capturing the network requests to AI services can cover any device on the same network, providing an integrated, cross-platform usage trace~\cite{swain2023leveraging}. Prior researchers argue that harnessing logs from existing networks is far easier to scale than deploying custom sensors on every user or location, including being expensive. But a managed network offers an “easily accessible data source” covering many users over long periods. This exemplifies how network traffic data can serve as a passive, ubiquitous sensor, aggregating activity across diverse contexts.

However, tapping network traces for research requires careful consideration of privacy and ethics. Internet traffic can reveal sensitive information about users’ activities, so data collection and analysis must be conducted anonymously. One solution is data minimization, collect only the data needed and at the lowest fidelity necessary~\cite{velykoivanenko2021those}. Researchers should explain what is being collected and how it will be used, and allow individuals to opt out or anonymize certain aspects. Techniques like aggregation, hashing identifiers, or focusing on metadata (e.g., timestamps, tool type) can help maintain user privacy while still yielding valuable behavioral insights. 

\section{Limitations \& Future Works}
Our privacy-first study design shaped what we could observe and decided what type of data we collected. We analyzed device-level traces (anonymized IPs; participants could register up to two devices) and did not link logs to identifiable individuals or message content. This choice protects participants but limits person-level triangulation with surveys or coursework data and prevents intent-level interpretation (e.g., why a student used a tool at a given moment). We also inferred AI activity from curated domains (e.g., \texttt{chatgpt.com}, \texttt{perplexity.ai}, Gemini-related endpoints, including google.com when summaries are integrated). This approach can over-attribute activity to AI when platforms embed models into routine services (e.g., search), and it can under-attribute activity when tools tunnel through unexpected endpoints or CDNs. Our sessionization rule (10-minute inactivity gap) makes patterns comparable across tools, but different thresholds would shift long-tail counts; we did not preregister alternatives or run a sensitivity analysis.

Coverage and reliability also varied. Students onboarded on staggered dates, some devices dropped offline, and a mid-study server outage created gaps that complicate longitudinal comparisons and likely inflated effort/frustration for a subset of participants. Although we observed consistent diurnal and finals-week spikes, the sample reflects a single university with many CS/CE graduate students and (for iOS) an optional App Privacy Report, which limits generalizability. Moreover, the cohort included students from a cybersecurity course, who were likely primed to think critically about privacy and may have been more sensitive to monitoring than a general student population. The cohort also came from a highly selective R1 university and was relatively young and tech-savvy; the acceptability and usability of such a system remain uncertain in other populations, such as older adults or participants in less technology-intensive contexts. Incentive structures (course credit) supported engagement but may have encouraged compliance behaviors (e.g., frequent VPN checks) that shaped usage rhythms.

Our study shows the promise of network-traffic-based ethnography, but several directions remain for strengthening its accuracy and interpretation.
One challenge is the lack of direct “ground truths.” Traditional ethnographic work provides researchers with firsthand observation, whereas our method relies on indirect interpretation. Future research could integrate complementary measures, such as ecological momentary assessments (EMAs), to capture activities in real-time and provide a richer baseline for validating traffic-based analyses~\cite{shiffman2008ecological}. Another direction is improving activity detection. Apple’s App Privacy Reports (APRs) indicate which apps participants use, but they do not reveal which features or functions are engaged. Many domains of interest are also not covered in the current APRs. Addressing these gaps will require controlled testing of different applications to generate broader APR datasets and the development of models capable of distinguishing more specific behaviors. For instance, future studies might train machine learning models that differentiate between calling and texting within WhatsApp. Pursuing these steps will help move from broad traffic traces to more precise interpretations of everyday technology use, ultimately deepening the value of infrastructural ethnography.

\bibliographystyle{ACM-Reference-Format}
\bibliography{sample-base}

\appendix



\end{document}